\documentclass[preprint,english,aps,prb,floatfix]{revtex4}
\usepackage{amssymb,amsmath}
\usepackage{graphicx}

\makeatletter

\usepackage{babel}

\renewcommand\@biblabel[1]{(#1)}

\makeatother

\begin{document}

\title{Effect of Bilayer Thickness on Membrane Bending Rigidity }

\author{H. Berm\'{u}dez}

%\email{bermudez@seas.upenn.edu}
\affiliation{Department of Chemical and Biomolecular Engineering, University of
Pennsylvania, Philadelphia, Pennsylvania 19104}

\author{D. A. Hammer}

\affiliation{Department of Chemical and Biomolecular Engineering, University of
Pennsylvania, Philadelphia, Pennsylvania 19104}

\author{D. E. Discher}

\affiliation{Department of Chemical and Biomolecular Engineering, University of
Pennsylvania, Philadelphia, Pennsylvania 19104}

\begin{abstract}

The bending rigidity $k_c$ of bilayer vesicles self-assembled from amphiphilic diblock copolymers has been measured using single and dual-micropipet techniques.  These copolymers are nearly a factor of 5 greater in hydrophobic membrane thickness $d$ than their lipid counterparts, and an order of magnitude larger in molecular weight $\bar{M}_n$.  The macromolecular structure of these amphiphiles lends insight into and extends relationships for traditional surfactant behavior.  We find the scaling of $k_c$ with thickness to be nearly quadratic, in agreement with existing theories for bilayer membranes.  The results here are key to understanding and designing soft interfaces 
such as biomembrane mimetics.

\end{abstract}

%\pacs{82.70.Uv, 68.65.-k, 87.68.+z}

\maketitle

%%%%%%%%
\section{Introduction}
%%%%%%%%

Thin films of surfactants are found in numerous contexts, ranging from emulsions and colloids to biological membranes. \cite{bsoft}  The bending rigidity of such films is a key determinant of many structures and processes including cell shape, fusion, and adhesion.  Surfactant studies to date have been largely limited in scope by synthetic chemistry capabilities, and thus, generalizations of emergent properties have been similarly constrained.  Several groups have nevertheless developed theoretical and numerical methods to predict properties of monolayers and bilayers from both continuum and molecular perspectives. \cite{bbloom,bigal,blebowski,bevans,bhelfrich}  Experimentally, the advent of techniques such as living anionic polymerization \cite{bpolim} has provided the means to create relatively monodisperse amphilphilic diblock copolymers -- the macromolecular analogues to ``short'' chain surfactants.  Besides finding novel uses in the above contexts, \cite{balexan} these macromolecules also serve to test the limits of existing theories.

In this letter, we examine the dependence of the bending rigidity $k_{c}$ on the hydrophobic thickness $d$ of closed bilayer membranes (vesicles).  Other important properties that vary with $d$, such as permeability and elasticity, are discussed elsewhere. \cite{bdischerfamily,bharry}  From the simplest models, a bilayer can be pictured as being composed of thin elastic shell(s).  The deformation of such a shell \cite{blandl,bbloom} is given by the 2-D Lame coefficients $\mu$ and $\lambda$ and the shell thickness $l$.  It follows that the area elastic modulus $K_A = \lambda + \mu$ and the bending rigidity $k_c = (\lambda + 2 \mu) l^2 / 12$.  Taking the shear modulus $\mu = 0$ for fluid membranes leads to the often-cited result, \cite{bevans,bhelfrich} 

\begin{equation}
k_{c}= \beta K_A d^{2}
\label{eq:bbend}
\end{equation}

(where $\beta$ is a constant).  This relation serves as motivation for our study.  Any possible interdigitation in the bilayer does not affect the scaling 
of $k_{c}$, but is instead reflected in the prefactor $\beta$. 
\cite{blandl,bbloom}  When the membrane is taken as uncoupled monolayers free to slide past one another, $\beta = 1/48$, and when taken as completely coupled monolayers, $\beta = 1/12$.  More complex descriptions \cite{bigal} include effects such as chain flexibility and associated entropy, with $k_{c}\sim d^{2.5}$.  Note that for a rigid plate, \cite{bboal} $k_{c}\sim d^{3}$, and 
thus we expect a strong scaling dependence, regardless of the exact details.

%%%%%%%%%%%%%%%%%%%%%%%% FIGURE %%%%%%%%%%%%%%%%%%%%%%%%%%%%%%%%%
\begin{figure}[!htbp]
\begin{center}
\includegraphics[width=3.0in]{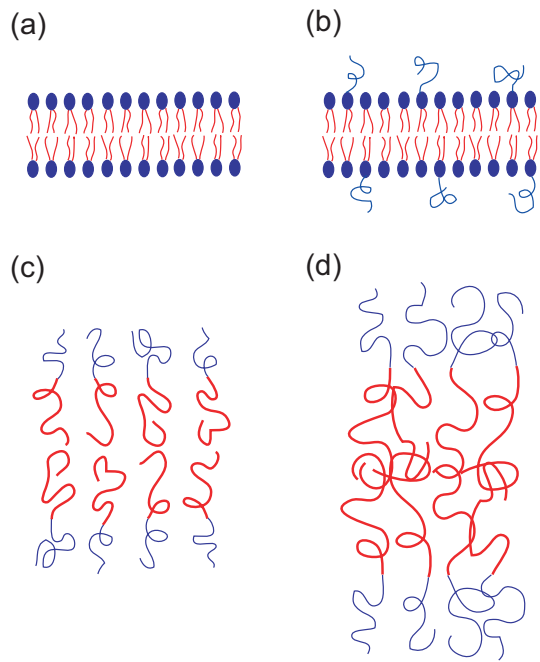}
\end{center}
\caption[Schematic of different bilayer membrane configurations.]
{\label{fig:bendtoon}
Schematic of different bilayer membrane configurations.
(a) Typical phospholipid membrane of hydrophobic thickness $d \approx 3$ nm.
(b) PEO-conjugated lipids can typically be accommodated only up to
15 mol \% before micellization begins.  At this low grafting density,
PEO is in the mushroom or marginal brush regime. \protect\cite{bdegennes}
(c) Diblock copolymer membranes are much thicker ($d > 8$ nm), and the PEO
is expected to be near or in the brush regime. \protect\cite{bwonbrush}
(d) At still larger $d$, chains may become entangled either laterally
within a monolayer or across monolayers.
}
\end{figure}
%%%%%%%%%%%%%%%%%%%%%%%% END FIGURE %%%%%%%%%%%%%%%%%%%%%%%%%%%%%

The picture of a biological membrane is clearly more complicated, as
it is perforated by integral proteins and has associated proteins
attached to its surfaces.  In particular, the presence of a surface 
brushy layer, or glycocalyx, is likely to contribute to bending 
resistance, as well as other properties and processes. \cite{bhairy}
Brushes on lipid membranes have been studied to a limited extent by 
attaching poly(ethylene oxide) (PEO) to phospholipid headgroups.  
These ``Stealth'' vesicles have found wide use in drug delivery 
applications, \cite{bstealth} and they presumably suppress immune 
response by means of the surface steric stabilization imparted by PEO.
However, lipid vesicles generally cannot accommodate more 
than $\approx 15$ mol \% of PEO-conjugated lipid due to resulting 
curvature effects and subsequent micellization. \cite{bdestab,bisrael}
This limitation on the amount of PEO that can be incoporated into lipid
membranes implies that the PEO is probably in a mushroom or marginal 
brush configuration at best. \cite{bdegennes}

In contrast to lipid membranes (and perhaps closer to biomembranes), 
polymer membranes have a dense PEO layer, likely in a brush or 
partially collapsed brush state. \cite{bwonbrush}  The effect of 
brushes on membrane elasticity and rigidity has been studied 
theoretically by several groups, \cite{blaradji,bmarques} but 
experimental verifications with lipid-based systems \cite{bevansfuzzy} 
are limited for the reasons mentioned above.
Various model scenarios are depicted schematically in 
Figure~\ref{fig:bendtoon}.
While it may not be possible to experimentally decouple the 
contributions of the hydrophobic and hydrophilic segments, direct 
measurements over a broad range of $d$ can still yield insight 
into general membrane behavior.

%%%%%%%%%%%%%%%%%%%%%%%% TABLE %%%%%%%%%%%%%%%%%%%%%%%%%%%%%%%%%
%\begingroup
%\squeezetable
\begin{table}[!htbp]
\caption[Structural details of vesicle-forming amphiphiles.]
{\label{bendtable}
Structural details of vesicle-forming amphiphiles.
The common biomembrane lipids SOPC (1-stearoyl-2-oleoyl
phosphatidylcholine) and DMPC (1,2-dimyristoyl phosphatidylcholine)
are included for comparison.
The hydrophilic fraction $f \approx 0.3-0.4$ is consistent
with lamellar structures.  Cryo-TEM provides direct measures of the
hydrophobic thickness $d$. \protect\cite{bharry}
%Asterisks denote literature values
%\cite{bdischerfamily,braw}.
}
\begin{center}\begin{tabular}{ccccc}
\hline
Amphiphile & Polymer & $\bar{M}_{n}$ & $~f~$ & $~~d~~$ \\
  & Formula & (kg/mol) &  & (nm)  \\
\hline
DMPC & --- & $0.68$ & $~\approx0.36~$ & $2.5$  \\
SOPC & --- & $0.79$ & $~\approx0.31~$ & $3.0$  \\
\textbf{OE7} & PEO$_{40}$-PEE$_{37}$ & $3.9$ & $0.39$ & $8.0$  \\
\textbf{OB2} & PEO$_{26}$-PBD$_{46}$ & $3.6$ & $0.28$ & $9.6$  \\
\textbf{OB18}& PEO$_{80}$-PBD$_{125}$& $10.4$ & $0.29$ & $14.8$ \\
\hline
\end{tabular}\end{center}
\end{table}
%\endgroup
%%%%%%%%%%%%%%%%%%%%%%%% END TABLE %%%%%%%%%%%%%%%%%%%%%%%%%%%%%%%%%

%%%%%%%%
\section{Experimental Methods}
%%%%%%%%
The polymers of PEO-polybutadiene (PEO-PBD) as well as 
PEO-poly(ethylethylene) (PEO-PEE) were synthesized by standard living 
anionic polymerization techniques. \cite{bhillmyer}
The number of monomer units in each block was determined by $^{1}$H NMR.
Gel permeation chromatography with poly\-sty\-rene standards was
used to determine number-average molecular weights $\bar{M}_{n}$
as well as polydispersity indices (always $<$ 1.10) (Table~\ref{bendtable}).  
The hydrophobic membrane thickness $d$ was previously determined by 
direct imaging of vitrified samples with cryogenic transmission electron microscopy (cryo-TEM). \cite{bharry}  Measurements of related micelles via cryo-TEM agree well with independent small-angle neutron scattering (SANS) results. \cite{bwonbrush,bwonscience}

Giant vesicles were prepared by typical film rehydration
techniques \cite{bbnb} and imaged under bright-field optics 
to provide distinct imaging of the vesicle membrane.
Narishige manipulators were connected to a custom manometer
system with pressure transducers (Validyne, Northridge, CA) for control
and monitoring of the aspiration pressure.  To obtain small pressures 
($\approx 10$ Pa), Mitutoyo digital micrometers were used to displace 
the relative heights of water in the manometer system. 
In the micromanipulation technique, \cite{bskalak,braw} a giant vesicle is
made slightly flaccid and then aspirated into a micropipet. From 
vesicle geometry, the applied pressure, and the aspirated projection 
length, one can calculate the imposed membrane tension $\Sigma $ and the 
relative area dilation $\alpha \equiv \Delta A/A_{o}$.   The area 
dilation is related to the membrane elastic constants, and is generally 
written \cite{bnuovo,bmilner} as a superposition of the entropic elasticity 
of surface undulations and direct membrane stretching against cohesive forces: 
\begin{equation}
\alpha = (k_BT / 8 \pi k_c) \ln(1 + c A \Sigma) + \Sigma / K_A,
\label{eq:balpha} 
\end{equation}
where the coefficient $c$ ($\approx 0.1$) depends on the type of mode 
expansion used to describe the undulations (\textit{e.g.,} plane wave or 
quasispherical approach).  From Eq.~(\ref{eq:balpha}), $k_c$ is directly
obtained from a plot of $\ln(\Sigma)$ versus $\alpha$.

\begin{figure}[!htbp]
\begin{center}
\includegraphics[width=2.5in]{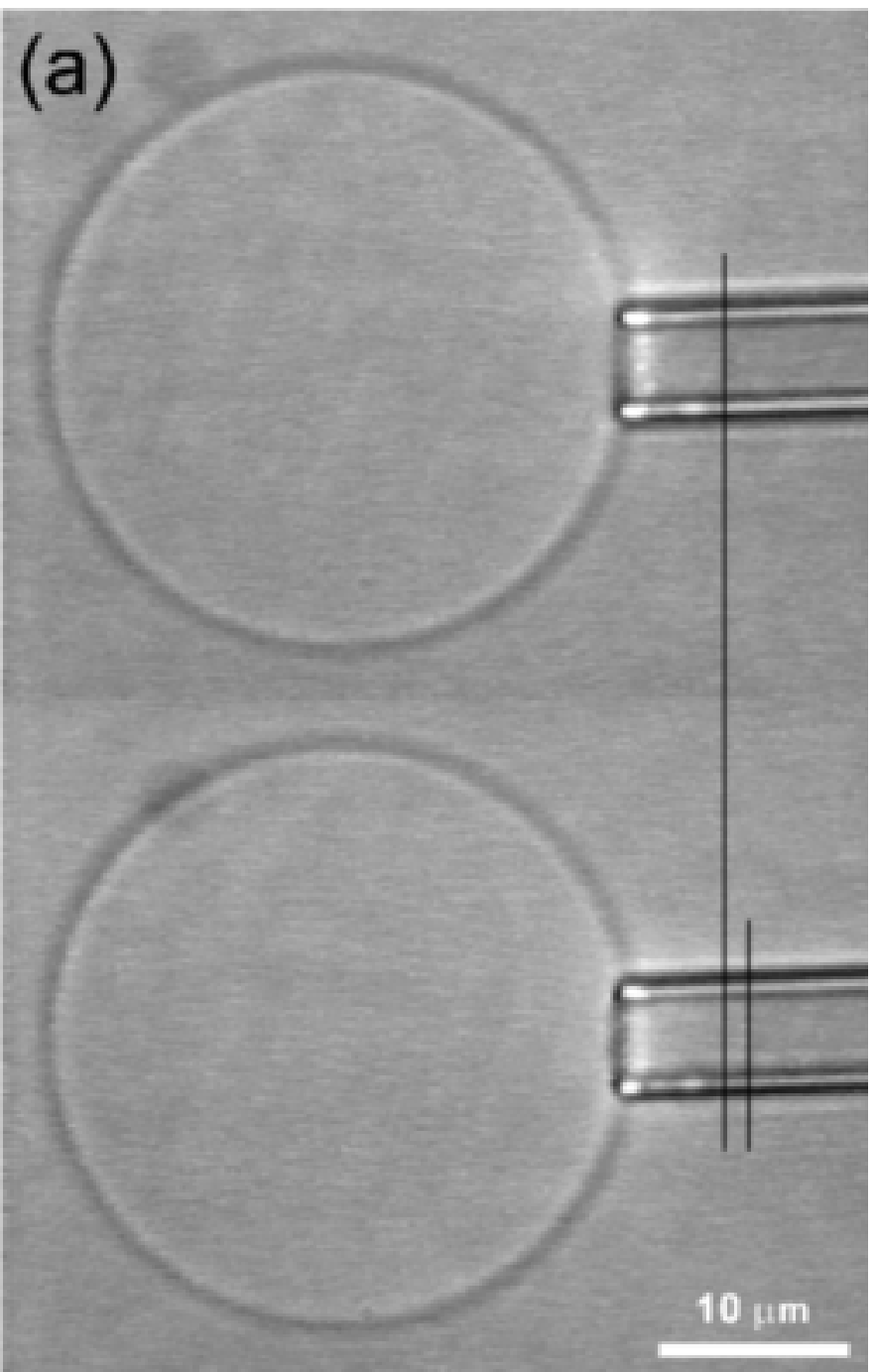}
\includegraphics[width=3.2in]{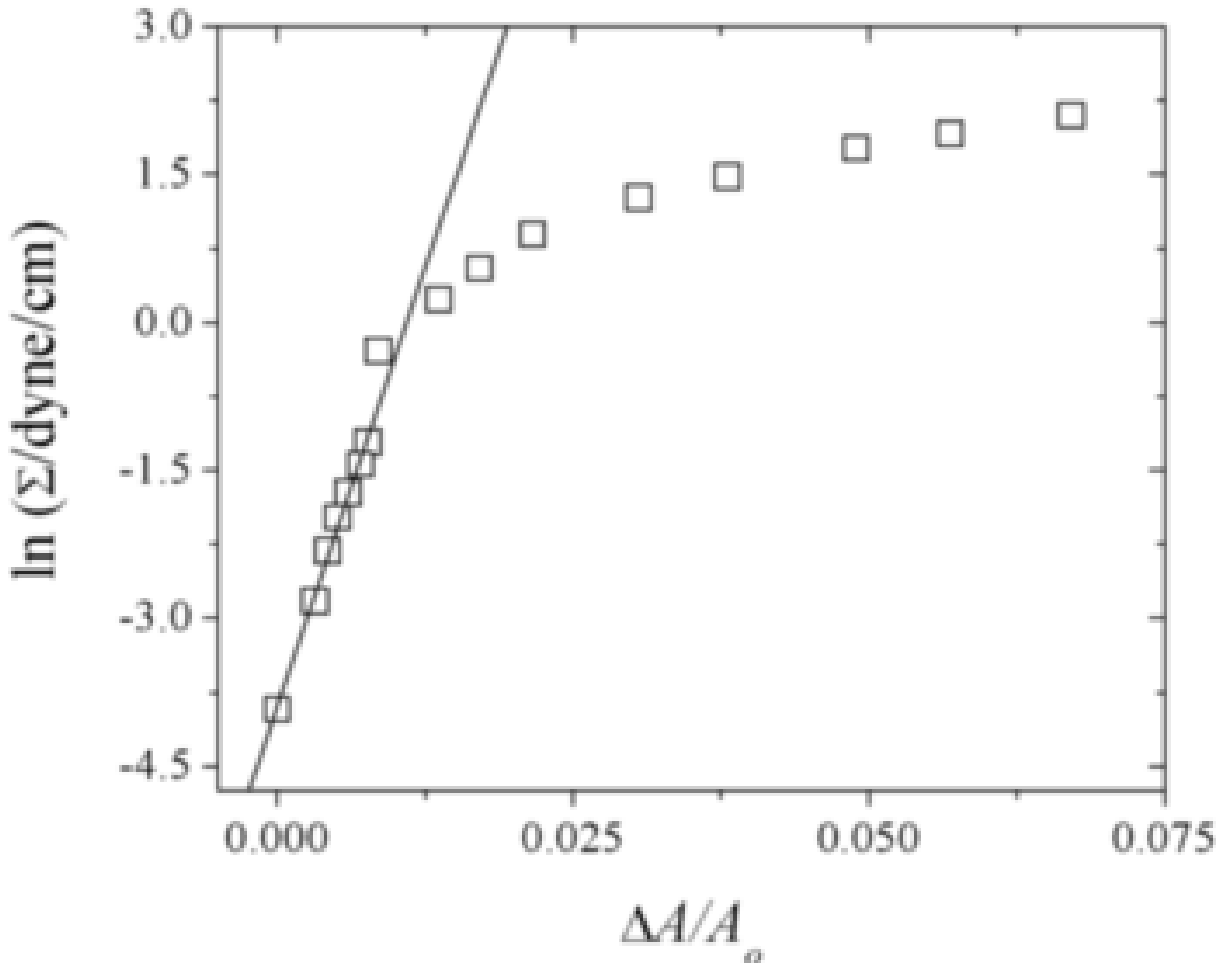}
\end{center}
\caption[Single pipet measurements of the bending rigidity $k_c$ (\textit{e.g.}, \textbf{OB2}).]
{\label{fig:bsingle}
Single pipet measurements of the bending rigidity $k_c$ 
(\textit{e.g.}, \textbf{OB2}).  
(a) Micropipet aspiration suppresses the membrane fluctuations at 
low tensions, smoothing out excess area in the membrane and resulting in an increased projection length $\Delta L$.
\protect\cite{bevanspip}  
(b) Corresponding plot of imposed tension vs. area dilation to determine 
$k_c$.  The slope is $8 \pi k_c / k_B T$ from Eq.~(\ref{eq:balpha}).
}
\end{figure}

Here we use two different, but related, micropipet aspiration 
techniques.  Soft membranes, those with $k_{c}$ about $10-100 k_{B}T$, 
will exhibit commensurate surface undulations. 
These fluctuations can be analyzed optically \cite{boptical} or 
be suppressed with a micropipet \cite{bevanspip} (Fig.~\ref{fig:bsingle}a).
We have chosen the latter approach, since it also gives other material
properties such as interfacial elasticity in a single measurement.
For stiffer membranes ($k_{c}>100k_{B}T$) thermal fluctuations are
suboptical and potentially dampened by viscous dissipation within
the bilayer, requiring other methods to determine $k_{c}$.
One particular micropipet approach was developed by Zhelev and co-workers 
to study neutrophils, \cite{bdoncho} which are quite stiff because of their 
highly viscous cytoplasmic interiors. Initial observations of polymer vesicles
formed from \textbf{OB18} (PEO$_{80}$-PBD$_{125}$) and higher $\bar{M}_n$ polymers indicated that viscous  effects were not negligible, \cite{bharry}
and thus would be better suited for this dual-pipet technique.  
By simultaneously aspirating a vesicle with two pipets (Fig.~\ref{fig:bdual}a), and accounting for the energy of the deformed regions, one obtains: 

\begin{equation}
\Delta P_s = f_0 k_c/R^{3}_{ps} + f_1 \Delta P_l / R_{ps}
\label{eq:bdoncho}
\end{equation}

where the coefficients $f_0$ and $f_1$ are functions of the pipet
and vesicle geometry. \cite{bdoncho} 
Aspiration pressures and pipet radii are denoted by $\Delta P$ 
and $R_p$, respectively, and the subscripts $l$ and $s$ distinguish 
between large and small pipets.
From Eq.~(\ref{eq:bdoncho}), $k_c$ is obtained from the intercept of 
$\Delta P_s$ vs. $\Delta P_l$.
%(Fig. \ref{fig:bdual}). 

%%%%%%%%
\section{Results and Discussion}
%%%%%%%%

The use of different techniques is necessitated by the strong effect
of $d$ on membrane properties.
Although the present study of diblocks is still limited in size, it 
considerably extends the range of membrane thickness over prior 
studies \cite{braw} with lipid vesicles ($d = 2.4 - 3.4$ nm).
A further difficulty with lipid or ``short'' chain systems is the 
ambiguity regarding the applicability of polymer theory.  
Polymeric systems are able to mitigate both of these limitations 
by virtue of their macromolecular nature.

\begin{figure}[!htbp]
% up and down
\begin{center}
\includegraphics[width=2.5in]{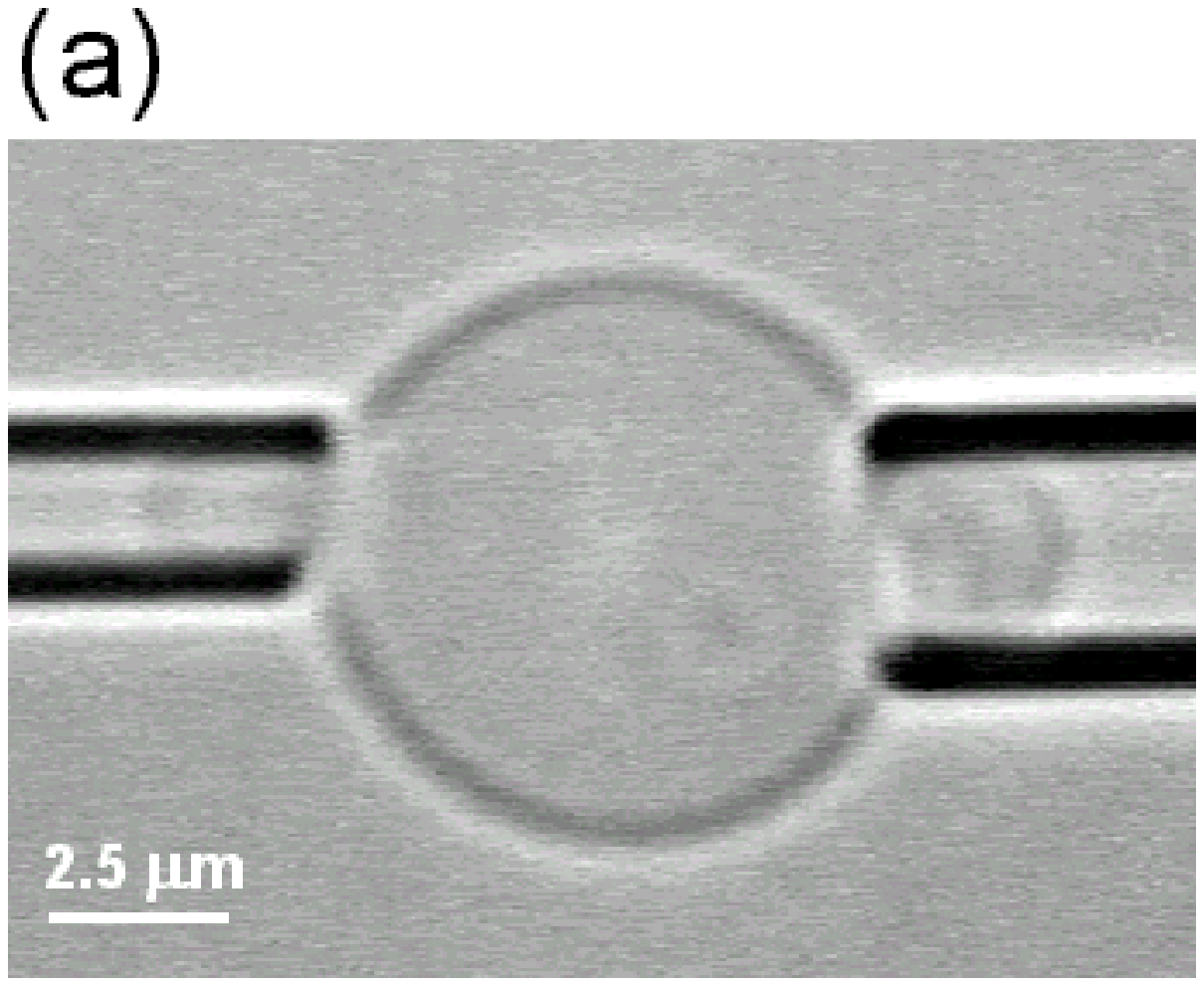}
\includegraphics[width=3.2in]{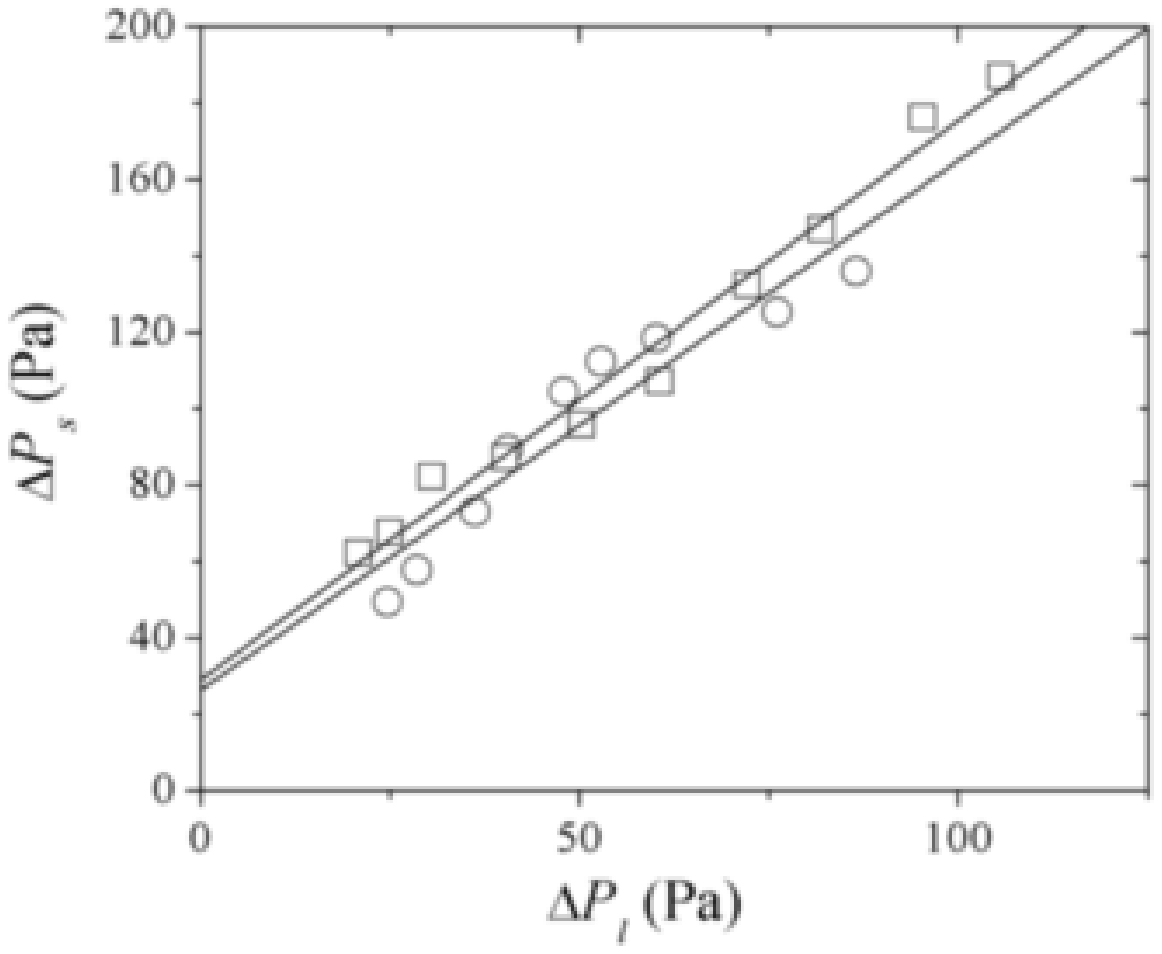}
\end{center}
% side by side
%\centering
%\begin{minipage}[c]{0.5\textwidth}
%\centering \includegraphics[width=3.2in]{single.ps}
%\end{minipage}%
%\begin{minipage}[c]{0.5\textwidth}
%\centering \includegraphics[width=3.2in]{Figures/bend/dual.eps}
%\end{minipage}
\caption[Dual pipet measurements of the bending rigidity $k_c$ (\textit{e.g.}, \textbf{OB18}).]
{\label{fig:bdual}
Dual pipet measurement of the bending rigidity $k_c$ (\textit{e.g.}, \textbf{OB18}).  
(a) A vesicle is partially aspirated into two pipets with different 
radii ($R_{pl}$ and $R_{ps}$).  
The larger pipet suction $\Delta P_l$
is used to take up excess area, while the smaller pipet aspiration 
pressure $\Delta P_s$, deforms a small region of the membrane.
(b) Corresponding aspiration curves for two different vesicles.  The
thermodynamic analysis of Zhelev \textit{et al.}, \protect\cite{bdoncho} 
leading to Eq.~(\ref{eq:bdoncho}), is used to extract $k_c$.
}
\end{figure}

Observations of vesicles made from \textbf{OE7} (PEO$_{40}$-PEE$_{37}$) and \textbf{OB2} (PEO$_{26}$-PBD$_{46}$) indicated that these membranes were 
relatively lipid-like and hence would have fairly low values of $k_c$.  
It is important to note that microscopic fluctuations persist at all
levels of tension, thereby renormalizing the area elastic modulus. 
\cite{bevanspip}  By virtue of their increased thickness, polymer 
membranes have larger $k_{c}$ and maximum areal strains $\alpha _{c}$,
\cite{bharry} and thus the correction to $K_A$ is less than $5 \%$.
For thinner and softer membranes, the apparent elastic modulus 
$K_{app} \equiv \partial\Sigma / (\partial A / A)$ is
related to the true modulus $K_A$ by
\begin{equation}
K_A/K_{app} = 1 + K_A k_BT / 8 \pi k_c \Sigma .
\label{eq:brenorm}
\end{equation}
From Eq.~(\ref{eq:brenorm}), the crossover tension $\Sigma_x$ between 
the regime dominated by fluctuations and the regime dominated by 
direct expansion is expected to decrease with $d$, since 
$\Sigma_{x} = K_A k_BT / 8 \pi k_c \sim d^{-2}$.
Indeed we find for \textbf{OB2} a lower value of $\Sigma_x$ compared to 
lipids with $k_c = 24.7 \pm 11.1 ~ k_BT$ ($N=6$).  
As a control, we also examined the prototypical lipid SOPC 
(1-stearoyl-2-oleoyl phosphatidylcholine) having $d = 3$ nm, 
and found $k_c = 16.4 \pm 6.9 ~ k_BT$ ($N=9$), in good agreement 
with published values. \cite{braw}
For the thickest membrane studied here, \textbf{OB18}, measurements 
with the dual pipet technique give $k_c = 466 \pm 157 ~ k_BT$ ($N=4$), 
substantially larger than any lipid 
(Table~\ref{proptable} and Fig.~\ref{fig:bdual}).

%%%%%%%%%%%%%%%%%%%%%%%% FIGURE %%%%%%%%%%%%%%%%%%%%%%%%%%%%%%%%%
%%%%%%%%%%%%%%%%%%%%%%%% END FIGURE %%%%%%%%%%%%%%%%%%%%%%%%%%%%%

%%%%%%%%%%%%%%%%%%%%%%%% FIGURE %%%%%%%%%%%%%%%%%%%%%%%%%%%%%%%%%
%%%%%%%%%%%%%%%%%%%%%%%% END FIGURE %%%%%%%%%%%%%%%%%%%%%%%%%%%%%

At first glance it may seem surprising that \textbf{OB2} has a lower value
of $k_c$ than \textbf{OE7}, even though it has a thicker hydrophobic core
$d$.  However, the PEO contribution to the overall membrane 
thickness might also need to be considered. \cite{bevansfuzzy,bcastro} 
The contrast from PEO via cryo-TEM is rather limited, essentially prohibiting direct measurement of the corona by this technique.  We can however, make an estimate of this contribution from polymer theory.
Inspection of Table~\ref{bendtable} reveals that \textbf{OB2} has a 
shorter PEO chain than \textbf{OE7}.  Evidence for a brush or partially 
collapsed PEO brush in related diblock micelles \cite{bwonbrush} suggests 
that the PEO chains are stretched relative to a Gaussian state.
Given the large incompatibility between PEO and PBD, a conservative 
yet reasonable estimate of the PEO length would come from assuming these 
diblocks are in the strong segregation limit (SSL).  
In the SSL, the characteristic domain length scales with molecular 
weight as $R \sim N^{2/3}$, with the result that \textbf{OE7} should be
about $10\%$ thicker than \textbf{OB2}.
While this difference is modest, it may be sufficient to explain the data
given the strong dependence of $k_c$ with membrane thickness.

%%%%%%%%%%%%%%%%%%%%%%%% TABLE %%%%%%%%%%%%%%%%%%%%%%%%%%%%%%%%%
%\begingroup
%\squeezetable
\begin{table}[!htbp]
\caption[Material properties of polymer and lipid vesicles.]
{\label{proptable}
Material properties of polymer and lipid vesicles.
The ``coupling'' constant $\beta$ as obtained from Eq.~(\ref{eq:bbend}). 
$\ast$ Literature values for selected systems. \cite{bdischerfamily,braw}
}
\begin{center}\begin{tabular}{cccc}
\hline 
Amphiphile & $~~~~~~K_A~~~~~~$ & $k_c$ & $~~\beta$ \\
 & (N/m) & ($k_B T$) &   \\
\hline
DMPC* & $0.234$ & $13.3 \pm 1.4$ & $~~1/26$ \\
SOPC* & $0.235$ & $21.4 \pm 1.4$ & $~~1/24$ \\
SOPC & $0.203$ & $16.4 \pm 6.9$ & $~~1/27$ \\
\textbf{OE7}* & $0.120$ & $ 33.3 \pm 7.1$ & $~~1/55$ \\
\textbf{OB2} & $0.098$ & $24.7 \pm 11.1$ & $~~1/90$  \\
\textbf{OB18} & $0.109$ & $465.5 \pm 157$ & $~~1/13$ \\
\hline
\end{tabular}\end{center}
\end{table}
%\endgroup
%%%%%%%%%%%%%%%%%%%%%%%% END TABLE %%%%%%%%%%%%%%%%%%%%%%%%%%%%%%%%%

Single pipet measurements were not performed with \textbf{OB18} but 
it is expected that any such attempts would be frustrated by long 
response times and viscous dissipation within the membrane.  
For a $d = 15$ nm membrane, Eq.~(\ref{eq:brenorm}) predicts
$\Sigma_x \approx 1 \times 10^{-2}$ dyne/cm, indicating the difficulty 
in measuring $k_c$ by this technique.
Thicker membranes ($d > 15$ nm) are predisposed to viscous and entanglement
effects, \cite{bharry} and were therefore not pursued further.  Similarly, we did not carry out dual pipet measurements for our thinner membranes.  As has been previously noted by Zhelev \textit{et al.}, \cite{bdoncho} dual pipet measurements are not very sensitive and hence are best suited for much stiffer membranes where such errors can be tolerated.

\begin{figure}[!htbp]
\begin{center}
\includegraphics[width=3.5in]{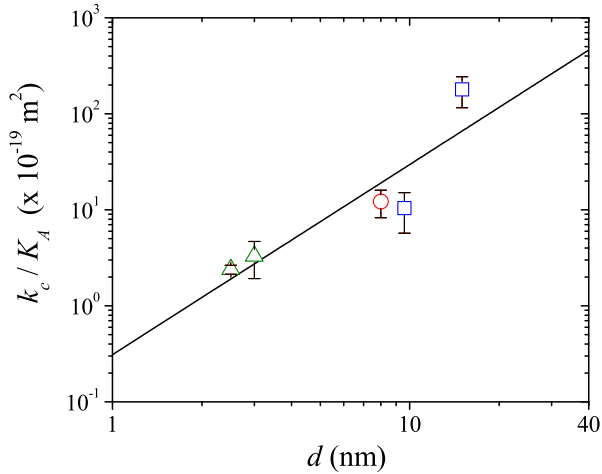}
\end{center}
\caption[Scaling of $k_c$ with membrane hydrophobic thickness $d$.]
{\label{fig:bscaling}
Scaling of $k_c$ with membrane hydrophobic thickness $d$.
Data are shown for DMPC ($\triangle$), \protect\cite{braw} 
SOPC ($\triangle$), \textbf{OE7} ($\circ$), \protect\cite{bdischerfamily}
\textbf{OB2} ($\Box $) and \textbf{OB18} ($\Box $) vesicles.  
\textbf{OB18} data are obtained by dual-pipet measurements, whereas 
all other data come from single-pipet measurements.  Line is
a least-squares fit, giving a scaling exponent of 2.0 ($R^2=0.809$) as 
predicted by Eq.~(\ref{eq:bbend}).  
}
\end{figure}

Combining the above results in Figure~\ref{fig:bscaling} shows good 
agreement with mean-field predictions for $k_c$, even though these 
systems are quite different from the classic phospholipid and surfactant 
systems motivating these theories.  
Rescaling these chemically distinct amphiphiles by the elastic area 
modulus $K_A$ is essential: for fluid membranes (such as these \cite{blee}), it has been shown both theoretically \cite{bhelf} and experimentally \cite{bharry} that $K_A$ is controlled by details 
of the interfacial chemistry and \textit{not} by the thickness $d$.  The extent of interdigitation is described by the numerical prefactor $\beta$, which 
in principle could vary with $d$, amphiphile persistence length, or 
chemistry.  Thus it is not so surprising that SOPC and \textbf{OB2} have
low values of $\beta$, given their visibly soft nature (Table~\ref{proptable}).  
For \textbf{OB18}, Eq.~(\ref{eq:bbend}) gives $\beta = 1/13$, implying 
that its leaflets are essentially coupled together.   This result 
is consistent with other observations that \textbf{OB18} is a highly 
viscous, and possibly entangled, membrane. \cite{blee}
%Simply fitting a common value of $\beta = 1/5$ from 
%Figure~\ref{fig:bscaling} is misleading and incorrect.  
Rigorously, both $\beta$ \textit{and} $K_A$ are needed to determine the
scaling, but we do not have independent measurements of the former.  
It is interesting to note that the product $\beta K_A$ may remain constant 
by having leaflet coupling (structure) offset changes in the interfacial 
tension (chemistry). 

%%%%%%%%%%%%%%%%%%%%%%%% FIGURE %%%%%%%%%%%%%%%%%%%%%%%%%%%%%%%%%
%%%%%%%%%%%%%%%%%%%%%%%% END FIGURE %%%%%%%%%%%%%%%%%%%%%%%%%%%%%

%%%%%%%%
\section{Conclusions}
%%%%%%%%
Sensitive micropipet techniques are used to examine single vesicles
and extract the bending rigidity $k_c$ for membranes self-assembled from 
amphiphilic diblock copolymers.  The macromolecular nature of these
amphiphiles considerably broadens the range over which these systems can
be studied, even to the point where bulk effects arise.
We find the scaling of $k_c$ with thickness $d$ to be nearly quadratic,
in agreement with existing theories for bilayer membranes.  
The results will likely influence future work on extending surfactant 
assemblies, since bending is often a predominant mode of deformation in 
soft matter systems.
Future work with mixing of short and long chains (\textit{e.g.}, \textbf{OB2}
with \textbf{OB18}) is expected to cause a dramatic lowering of $k_c$ and
is yet another means of controlling interfacial properties.

%\begin{acknowledgments}
\section*{Acknowledgments}
The authors thank the Bates group at the University of Minnesota for 
synthesis and characterization of the copolymers in Table~\ref{bendtable}.  
Funding was provided by NSF-MRSEC and NASA. 
%\end{acknowledgments}

\end{document}